**Bartosz KAMIŃSKI[1], Agata ZIELIŃSKA[1], Anna MUSIAŁ[1], Ching-Wen SHIH[2], Imad LIMAME[2], Sven RODT[2], Stephan REITZENSTEIN[2], Grzegorz SĘK[1]**

Wrocław University of Science and Technology, Department of Experimental Physics (1), Technical University of Berlin, Institute of Solid State Physics (2)
ORCID: 1. 0009-0007-0640-6369; 2. 0000-0002-7568-5257; 3. 0000-0001-9602-8929; 4. 0009-0003-6483-7862; 5. 0009-0003-2807-8424; 6. 0000-0001-6275-4965; 7. 0000-0002-1381-9838; 8. 0000-0001-7645-8243

# Optical characteristics of cavity structures with Al$_{0.2}$Ga$_{0.8}$As/Al$_{0.9}$Ga$_{0.1}$As distributed Bragg reflectors and In$_{0.37}$Ga$_{0.63}$As quantum dots as the active region

**Streszczenie.** Scharakteryzowano optycznie strukturę z planarną wnęką optyczną utworzoną przez zwierciadła Bragga z Al$_{0.2}$Ga$_{0.8}$As/Al$_{0.9}$Ga$_{0.1}$As oraz z kropkami kwantowymi In$_{0.37}$Ga$_{0.63}$As/GaAs jako obszarem aktywnym. Optymalizacja obszaru aktywnego wymaga zidentyfikowania głównych mechanizmów wygaszania fotoluminescencji. W celu zapewnienia wysokiego współczynnika sprzęgania emisji do modu wnęki, konieczne jest spełnienie warunku dopasowania spektralnego wnęki i emitera, co określono poprzez porównanie widm odbicia z wnęki optycznej i fotoluminescencji z obszaru aktywnego. Pomiar odbicia pozwolił na zweryfikowanie badanej struktury poprzez porównanie z symulacją. **(Właściwości optyczne wnęki optycznej utworzonej przez zwierciadła Bragga Al$_{0.2}$Ga$_{0.8}$As/Al$_{0.9}$Ga$_{0.1}$As z obszarem aktywnym w postaci kropek kwantowych In$_{0.37}$Ga$_{0.63}$As/GaAs)**

**Abstract.** We characterized optically a structure with Al$_{0.2}$Ga$_{0.8}$As/Al$_{0.9}$Ga$_{0.1}$As distributed Bragg reflector (DBR) based planar cavity with In$_{0.37}$Ga$_{0.63}$As/GaAs quantum dots (QDs) in the active region. Optimization of the active region requires identification of the main quenching mechanisms of photoluminescence from QDs. To maximize efficiency of QD emission coupling into the cavity mode, a spectral matching condition between the cavity and the emitter needs to be fulfilled. This was checked by comparing the reflectance spectrum of the cavity with the QD emission spectrum. Reflectivity spectra allowed us to verify the investigated structure by comparing them with simulations.

**Słowa kluczowe**: zwierciadło Bragga, kropka kwantowa, wnęka optyczna, lasery półprzewodnikowe, bliska podczerwień
**Keywords**: distributed Bragg reflector, quantum dot, optical cavity, semiconductor laser, near infrared spectral range

## Introduction

Despite the great progress made in the field of vertically emitting semiconductor microlasers within the last two decades, there is still a room for improvements when concerning their performance, especially when aiming at very specific applications, like e.g. laser-based sensing systems for particular substances. To exploit high performance silicon detectors in the near infrared range, GaAs-based lasers are advantageous. Optically pumped micropillar lasers realized in this material system suffer from low power conversion efficiency due to the strong optical pump power absorption in the top Al(Ga)As/GaAs DBR before it reaches the GaAs cavity and thus the active region [2]. A simple solution to the low power conversion efficiency problem has been proposed recently and relies on replacement of the conventional Al(Ga)As/GaAs DBRs with Al$_x$Ga$_{1-x}$As/Al$_y$Ga$_{1-y}$As DBRs to blueshift the absorption edge, when compared to GaAs [2],[3].

In this article we report on the optical characterization of building blocks of a laser structure with Al$_{0.2}$Ga$_{0.8}$As/Al$_{0.9}$Ga$_{0.1}$As DBR cavity and In$_{0.37}$Ga$_{0.63}$As quantum dots (QDs) active region for near infrared spectral range. Zero-dimensional structures offer good thermal stability of the threshold current and spectrally broad gain in the range of few tens of nanometers, at the expense of lower output power. The latter is not an issue in the case of gas detection applications where powers on the level of 100 μW can be sufficient and easily achievable with QDs. Spectrally broad gain enables to cover wider spectral range what gives a potential opportunity to fabricate a tunable laser or an array of lasers differing in output wavelength on a single wafer, both very much demanded in practical gas analyzer systems.

## Investigated structures

The investigated structures were grown by metal-organic chemical vapor deposition (MOCVD) on (001) GaAs substrate. GaAs planar resonator with a thickness of $\lambda$ was created between Al$_{0.9}$Ga$_{0.1}$As/Al$_{0.2}$Ga$_{0.8}$As DBRs - 33.5 pairs of layers in the bottom mirror and 29.5 pairs in the top mirror. The upper mirror is additionally covered with a layer of GaAs with a thickness of $\lambda/4$ to prevent oxidation of the aluminum-containing layers. Inside the cavity there is a single layer of In$_{0.37}$Ga$_{0.63}$As/GaAs QDs formed by self-assembly in Stranski-Krastanow growth mode. The dots are optimized to have a high surface density of approximately $1.4 \times 10^{10}$ cm$^{-2}$ to provide sufficient optical gain for laser applications. We also grew a reference structure without DBRs for optical characterization of the active region independent of the cavity. Both structures are schematically shown in Fig. 1.

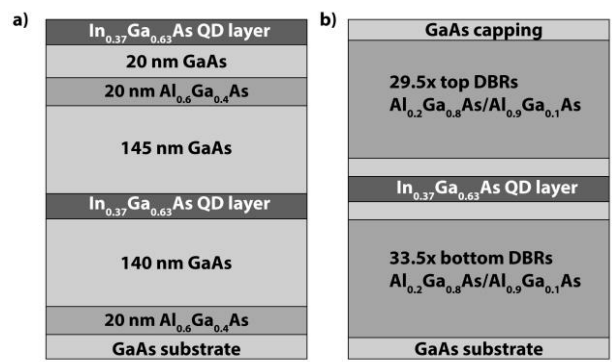

Fig. 1. Layer scheme of the a) reference sample (without optical cavity) and the b) sample with the full DBR-based cavity

## Experimental details

Optical characterization of both structures was carried out with spectroscopy methods such as photoluminescence (PL), temperature-dependent PL and reflectance measurements. The PL measurements were conducted with continuous wave and non-resonant excitation using semiconductor laser with 640 nm output wavelength. Laser beam was focused on the sample surface to a spot with 1 μm diameter using a long working distance infinity corrected microscope objective with numerical aperture of 0.4. The signal was analyzed with 1 m focal length monochromator and nitrogen cooled InGaAs multichannel linear array detector. For temperature-dependent measurements the sample was mounted in a continuous-flow liquid helium cryostat which enabled us to perform measurements in a wide temperature range from 10 K up to 400 K.

Reflectance measurements were conducted at room temperature (RT) with broadband light from a halogen lamp modulated at 280 Hz frequency with a mechanical chopper. The signal was further dispersed by 30 cm focal length monochromator providing 0.5 meV spectral resolution and detected with an InGaAs photodiode followed by a lock-in amplifier for phase sensitive measurements. To comply the spectral characteristics of the experimental setup, a reflectivity spectrum from a silver mirror was measured in the same experimental conditions.

**Calculations of reflectivity spectra**

In order to verify the nominal parameters of the DBRs, reflectance spectra of the full cavity (Fig. 1b) were simulated using the transfer matrix method [4]. Absorption of incident light in the structure is not taken into account in this model. It can be neglected in our case as the absorption edge of the DBR and cavity material is at higher energies than the DBR stopband. Input parameters for calculating the reflectance spectrum are thickness and refractive index (which we assumed constant in the considered spectral range and equal to the refractive index for the central wavelength of the stopband at room temperature) of all the layers in the structure. This method allows not only to obtain the reflectance spectrum in order to compare with the experimental data and verify the parameters of the fabricated structures.

**Results and discussion**

We measured temperature-dependent PL on the reference sample (Fig. 1a) to determine the spectral range of emission from QDs, and its thermal stability to identify the main PL quenching mechanisms.

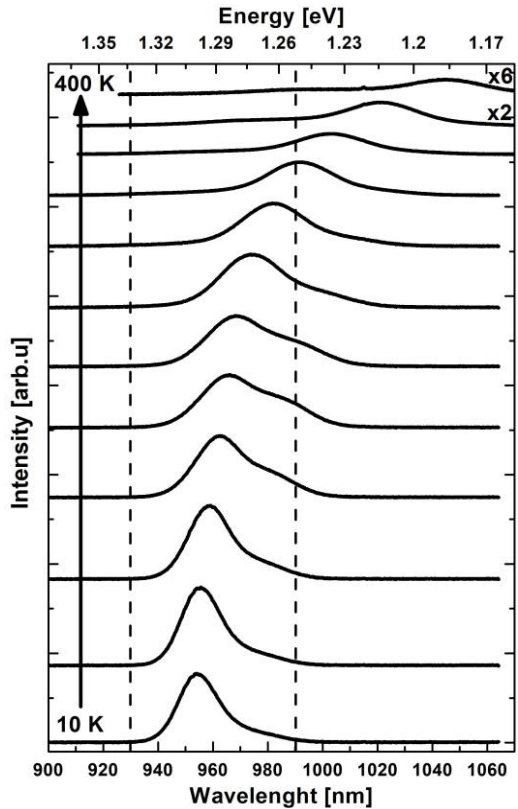

Fig. 2. PL as a function of temperature measured with ME for the reference sample with QDs (Fig. 1a). Dashed lines indicate the range of maximum reflectance of the DBR-based cavity (Fig. 1b)

The measurements were carried out with low excitation power density (LE - 350 W/cm$^2$) and medium excitation power density (ME - 3.5 kW/cm$^2$). It can be noticed that the PL spectrum is rather complex, but in further analysis we consider only the highest intensity emission band corresponding to emission from the ensemble of QDs. It can be observed for the whole temperature range up to 400 K. At low temperatures (LT) it is spectrally located within the stopband range of the DBR-based cavity as can be seen in Figure 2. The obtained PL spectra were fitted using Gaussian function what enabled us to acquire parameters such as peak position, integrated intensity and full width at half maximum (FWHM).

The PL of QDs shifts towards higher energies with increasing temperature (Fig. 3) approximately following the empirical equation for the alloy composition and temperature dependence of the semiconductor bandgap in In$_x$Ga$_{1-x}$As [5]:

$$(1) \quad E_g(x,T) = 0.42 + 0.625(1-x) - \left(\frac{5.8}{T+300} - \frac{4.19}{T+271}\right) * 10^{-4}T^2(1-x) - \frac{4.19*10^{-4}}{T+271}T^2 + 0.475(1-x)^2$$

where the equation uses bandgap energy at 0 K and material Varshni parameters α and β of binary alloys InAs and GaAs. To achieve good agreement of experimental data with above equation we needed to add 0.3 eV with respect to the energy bandgap of bulk In$_{0.37}$Ga$_{0.63}$As. This difference origins from carrier confinement in QDs and the origin of the optical transitions which occurs between QD discrete states, in contrast to emission from band edges in bulk.

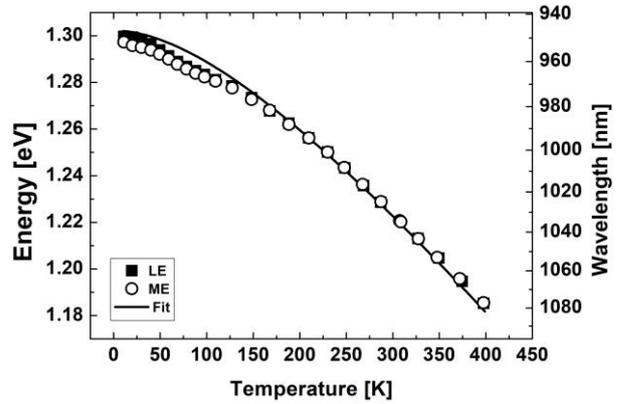

Fig. 3. PL QD peak position as a function of temperature for both excitation power densities fitted with empirical Eq. 1 (reference sample)

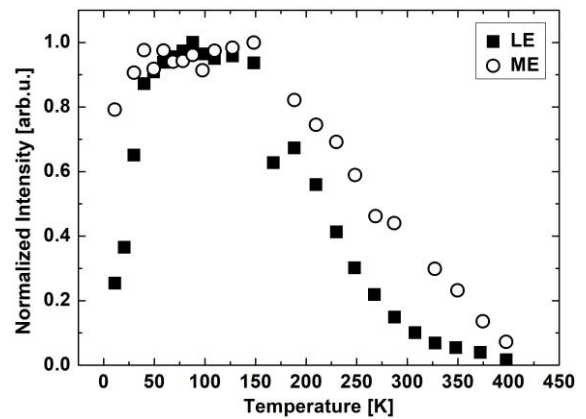

Fig. 4. Normalized integrated intensity of QD emission as a function of temperature for both excitation power densities (reference sample)

The PL intensity first increases rapidly until 40 K as can be seen in Figure 4. The amplitude of this increase is smaller for higher excitation power. This behavior indicates thermally activated transfer of carriers trapped by localized states in the dots surrounding (e.g. shallow defects) to QDs. The amount of transferred carriers increases up to the saturation level - filling all available states in the QDs. It can be observed for excitation powers below QD saturation condition. After this point, the PL intensity is almost constant up to 150 K where it starts to decrease as expected due to the increasing probability of non-radiative recombination and thermally-activated escape of carriers from the QDs due to finite depth of confining potential.

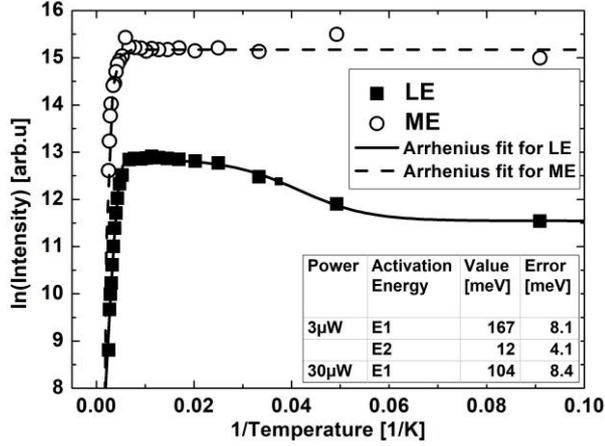

Fig. 5. Arrhenius plot for reference sample

Based on the determined integrated PL intensities we made Arrhenius plots (Fig. 5). The analysis of the temperature dependence of the QD emission intensity made it possible to determine the activation energies responsible for the PL quenching. For the ME case, we used standard empirical Arrhenius function with a single activation process [8]:

$$(2) \quad I(T) = \frac{I_0}{1 + B_1 \exp\left(-\frac{E_1}{k_B T}\right)}$$

where $I$ denotes the intensity of the PL, $I_0$ is the initial (maximum) intensity, $E_1$ is the characteristic activation energy and $B_1$ is a constant, i.e. an amplitude parameter reflecting efficiency of the given quenching process. For the case of LE there is observed an initial increase of the PL intensity with temperature, suggesting thermal activation of carriers from the traps, therefore we used a modified empirical Arrhenius function [9]:

$$(3) \quad I(T) = \frac{I_0 + I_P \left[1 - \left(1 - \frac{1}{1 + B_2 \exp\left(-\frac{E_2}{k_B T}\right)}\right)\right]}{1 + B_1 \exp\left(-\frac{E_1}{k_B T}\right)}$$

where parameters denote the same as above, except now we have to consider $E_1$ and $E_2$ as activation energies with the corresponding $B_1$ and $B_2$ constants. Smaller $E_2$ activation energy on the order of few meV (Table in the inset of Fig. 5) corresponds to the trap activation and subsequent QD filling, whereas the larger activation energy $E_1$ corresponds to the thermally activated loss of charge carriers. The coupling constant $B_1$ ($B_2$) describes the efficiency of intensity decrease (increase) with temperature. Here, we also have to take into account $I_p$ which is the reservoir-gained intensity to describe the initial increase of the emission intensity.

Comparison of the obtained activation energies (see the table in the inset of Fig. 5) with the band structure allows to identify the dominant escape channels of carriers from QDs. The characteristic PL quenching process energy exceeds 100 meV corresponding to carrier escape to the GaAs barrier. The difference between QD ground state transitions energy and GaAs bandgap is about 200 meV, which is further divided between the conduction and valence band according to band offsets (not necessarily equally), translating into about 100 meV needed for efficient escape of electrons or holes (or both) from the QD potential.

The inhomogeneous broadening of emission from a QD ensemble ranges from approx. 20 meV (at 10 K) to about 80 meV (at 290 K), similar for both the LE and ME excitation conditions. It translates into the FWHM of approx. 65 nm at room temeprature, which shows suitability of such active region for broadband gain and therefore the laser tunability.

The cavity characteristics like the mode wavelength, central wavelength of the DBR stopband and width of the stopband were obtained by conducting reflectance measurements on the full cavity structures with DBRs (see Fig. 6). To confirm the fabrication accuracy, we performed calculations with nominal parameters (top graph in Fig. 6). The refractive indices of the (Al)GaAs were estimated by extrapolation and interpolation from reported values [10-11].

Table 1. Parameters used in calculations via transfer matrix method

| Layers | Refractive index | Nominal layer thickness [nm] | Fitted layer thickness [nm] |
|---|---|---|---|
| GaAs capping | 3.5367 | 67.2 | 68.2 |
| GaAs cavity | 3.5367 | 268.0 | 272.9 |
| $Al_{0.2}Ga_{0.8}As$ | 3.4247 | 69.9 | 70.4 |
| $Al_{0.9}Ga_{0.1}As$ | 3.0291 | 79.7 | 79.6 |

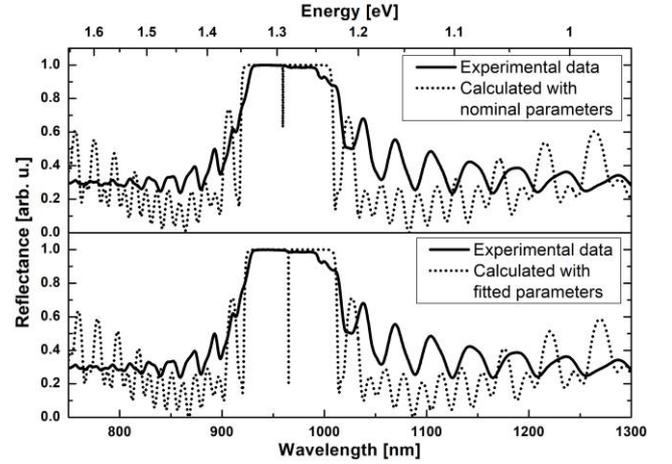

Fig. 6. Reflectance spectrum measured from the sample with DBRs compared with calculations with nominal parameters (top) and fitted parameters (bottom)

From comparison of the experimental and calculated reflectivity data (Fig. 6 top graph) it can be seen that experimental spectrum lacks cavity mode and the contrast of interference fringes beyond the stopband region is lower. The reason for this is low spectral resolution of used experimental setup and high Q-factor of the DBR-based cavity. To verify the fabrication accuracy we ran the calculations again to find out the real parameters of our sample, by slightly tuning the thickness of DBR layers from the nominal values until we achieved a good agreement of calculated spectrum with experimental data (see bottom part of Fig.6). It can be seen in the Table 1 with a summary of the used parameters in the simulations, that a deviation of layers thicknesses from nominal values is less than 1 nm for thin layers and less than 5 nm for GaAs cavity. This confirms overall good accuracy of the used growth method.

We verified that growth of the DBRs did not influence the active region significantly. The growth process of QDs is reproducible which we confirmed by measuring PL on the full cavity sample from the edge of the structure (directly exciting the active region and collecting mainly emission into the leaky modes – Fig. 7). The slight difference in QD emission between the two samples is because of modified strain conditions due to the existence of the bottom DBR, which then affects the QDs' morphology.

Finally, comparing emission from the active region in the full cavity structure (Fig. 1b) with reflectance of the cavity we can see that the PL peak falls within the stopband in the investigated range of temperatures, i.e. when tuned from LT to RT (Fig.7). The maximum emission peak should be aligned to the center of the maximum reflectance of the structure to ensure high QD-cavity coupling efficiency. Taking this into account the structure needs to be further optimized to fully exploit its potential. For RT applications QDs emission needs to be blueshifted to fit better the cavity reflectance maximum.

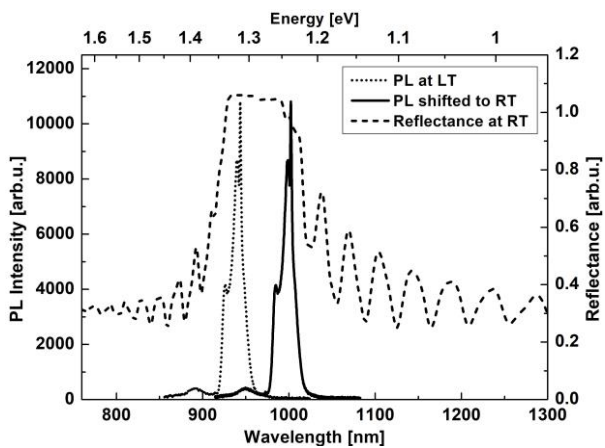

Fig. 7. Photoluminescence (from the edge) of the active region measured at 10 K (dotted line) and shifted to RT (solid line) in comparison to reflectance of the cavity at RT – dashed line (both measured from structure with DBRs)

**Conclusions**

Cavity structure with AlGaAs-based DBRs and InGaAs QDs as the emitters was characterized by optical spectroscopy, supported with modeling the reflectivity spectra. It is obtained that the LT emission of the active region fits well into the cavity stopband. The emission of the QDs must be slightly blueshifted if we consider the structure for RT applications. This can be achieved by decreasing the In content in the QD material. However, this will make the QD confining potential shallower and therefore, the GaAs barriers have to be replaced by wider bandgap material (e.g., AlGaAs) providing deeper confining potential for carriers to maintain efficient RT emission. The thermal stability of the emission has been verified by determining the activation energies of the PL quenching processes to exceed 100 meV corresponding to carrier escape to the GaAs barrier. The investigated In(Ga)As/GaAs QDs have a confining potential deep enough to emit at RT. On this basis, it can be concluded that they are potentially good for use as an active region in VCSELs and other optoelectronic devices in the near infrared spectral range. Absence of the cavity mode in the experimental reflectance spectrum due to low spectral resolution suggests that the cavity has a high Q factor suitable for laser applications. Indeed Q-factors exceeding 10 000 have been observed for such cavities [12]. The emission band width (FWHM) at RT is around 80 meV (65 nm), which is a good base for obtaining broad gain and hence application in tunable lasers.

*We sincerely appreciate all Reviewers valuable comments, which helped us in improving the quality of the article. We acknowledge the suggestion of conducting reflectance measurements at an angle to observe the cavity mode, which we will be happy to verify in subsequent measurements.*

*This work is supported by the joint Poland-Brandenburg Photonics Programme of the Investitionsbank Berlin and the Polish National Centre for Research and Development within the scope of project „QD-Sense: Cost-efficient gas sensing system based on wavelength tunable quantum-dot VCSEL arrays with nanogratings"*

**Authors:** *Bartosz Kamiński, Department of Experimental Physics, Wrocław University of Science and Technology, Wybrzeże Wyspiańskiego 27, 50-370 Wrocław, Poland, E-mail: 250201@student.pwr.edu.pl; Agata Zielińska, Department of Experimental Physics, Wrocław University of Science and Technology, Wybrzeże Wyspiańskiego 27, 50-370 Wrocław, Poland, E-mail: agata.zielinska@pwr.edu.pl; Dr Anna Musiał, Department of Experimental Physics, Wrocław University of Science and Technology, Wybrzeże Wyspiańskiego 27, 50-370 Wrocław, Poland, E-mail: anna.musial@pwr.edu.pl; Ching-Wen Shih, Institute of Solid State Physics, Technical University of Berlin, Hardenbergstraße 36, D-10623 Berlin, Germany, E-mail: ching-wen.shih@tu-berlin.de; Imad Limame, Institute of Solid State Physics, Technical University of Berlin, Hardenbergstraße 36, D-10623 Berlin, Germany, E-mail: imad.limame@tu-berlin.de; Dr Sven Rodt, Institute of Solid State Physics, Technical University of Berlin, Hardenbergstraße 36, D-10623 Berlin, Germany, E-mail: sven.rodt@tu-berlin.de; Prof. Dr Stephan Reitzenstein, Institute of Solid State Physics, Technical University of Berlin, Hardenbergstraße 36, D-10623 Berlin, Germany, E-mail: stephan.reitzenstein@tu-berlin.de; Prof. Dr Grzegorz Sęk, Department of Experimental Physics, Wrocław University of Science and Technology, Wybrzeże Wyspiańskiego 27, 50-370 Wrocław, Poland, E-mail: grzegorz.sek@pwr.edu.pl*